\documentclass[conference,9pt]{IEEEtran}
\IEEEoverridecommandlockouts

\usepackage{balance}
\usepackage{cite}
\usepackage{amsmath,amssymb,amsfonts}
\usepackage{algorithmic}
\usepackage{graphicx}
\usepackage{microtype}
\usepackage{textcomp}
\usepackage{xcolor}
\def\BibTeX{{\rm B\kern-.05em{\sc i\kern-.025em b}\kern-.08em
    T\kern-.1667em\lower.7ex\hbox{E}\kern-.125emX}}

\usepackage{subfig}
\usepackage{tabularx}
\usepackage{multirow}
\usepackage{booktabs}
\usepackage{siunitx}
\newcolumntype{d}[1]{D{.}{.}{#1}}

\newcommand{\mytable}{
	\centering
	\renewcommand{\arraystretch}{1.2}
}

\newcolumntype{L}{>{\raggedright\arraybackslash}X}
\usepackage[prependcaption,textsize=scriptsize]{todonotes}
\definecolor{mycolor}{HTML}{FF6600}

\begin{document}

\title{
Speech Recognition for Automatically Assessing \\ Afrikaans and isiXhosa Preschool Oral Narratives
\thanks{
We gratefully acknowledge the British Academy for their research grant (ECE 190079) and Fab Inc.\ for their research grant (W1/01B).
We would like to also thank Julian Herreilers, Retief Louw, Emma Sharratt, Luke Crowley and Shelley O'Carroll for helpful suggestions and assistance with the data.%
}%
}


\author{
    \IEEEauthorblockN{
         Christiaan Jacobs\IEEEauthorrefmark{1}, Annelien Smith\IEEEauthorrefmark{1}, Daleen Klop\IEEEauthorrefmark{1}, Ond\v{r}ej Klejch\IEEEauthorrefmark{2}, Febe de Wet\IEEEauthorrefmark{1}, Herman Kamper\IEEEauthorrefmark{1}
    }
    \IEEEauthorblockA{
        \IEEEauthorrefmark{1}\textit{Stellenbosch University, South Africa}\\
        \IEEEauthorrefmark{2}\textit{University of Edinburgh, United Kingdom}
    }
}

\maketitle

\begin{abstract}
We develop automatic speech recognition (ASR) systems for stories told by Afrikaans and isiXhosa preschool children. Oral narratives provide a way to assess children's language development before they learn to read.
We consider a range of prior child-speech ASR strategies to determine which is best suited to this unique setting. Using Whisper and only 5 minutes of transcribed in-domain child speech, we find that additional in-domain adult data (adult speech matching the story domain) provides the biggest improvement, especially when coupled with voice conversion. Semi-supervised learning also helps for both languages, while parameter-efficient fine-tuning helps on Afrikaans but not on isiXhosa (which is under-represented in the Whisper model). Few child-speech studies look at non-English data, and even fewer at the preschool ages of 4 and 5. Our work therefore represents a unique validation of a wide range of previous child-speech ASR strategies in an under-explored setting.
\end{abstract}

\section{Introduction}
\label{sec:intro}

Less than 20\% of South African 10-year olds can read for meaning~\cite{vanstaden2023pirls}. 
But problems often start much earlier.
Oral language skills provide the foundation for literacy and language development in preschool and predict later literacy and reading abilities~\cite{chiu2018simple,hjetland2020preschool,babayiugit2021linguistic}.
While literacy and reading are essential, these skills cannot be assessed in preschool children aged 4 to 5.
However, narrative and storytelling skills can be evaluated at this stage. Using narrative assessments to identify children at risk for literacy and reading difficulties early on could enable timely interventions, improving children's chances of later success~\cite{schick2010development,reese2010maternal,oakhill2012precursors,gardner2015oral}.
To make such assessments possible in overcrowded preschool classrooms, this current work forms part of an initiative to develop automated systems for assessing oral narratives of preschool children in South Africa.

\begin{figure}[!b]
\centering
	\subfloat{\includegraphics[width=0.329\linewidth]{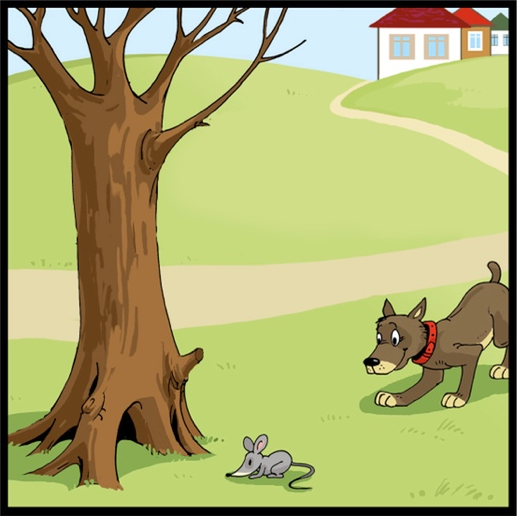}}\hfill
	\subfloat{\includegraphics[width=0.329\linewidth]{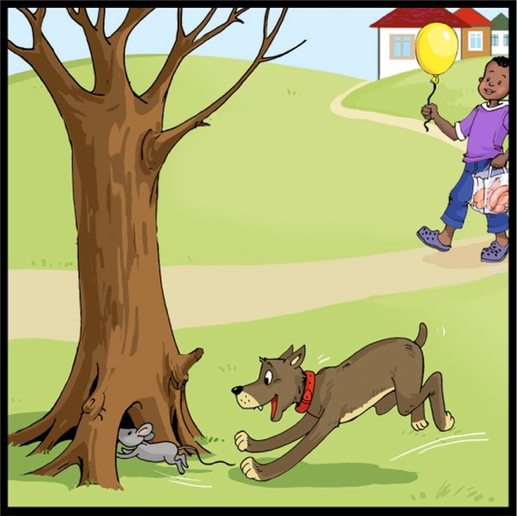}}\hfill
	\subfloat{\includegraphics[width=0.329\linewidth]{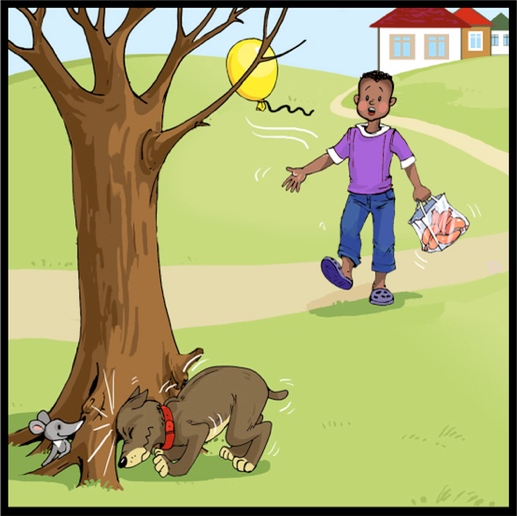}}
	
	\vspace{-3mm}
	
	\subfloat{\includegraphics[width=0.329\linewidth]{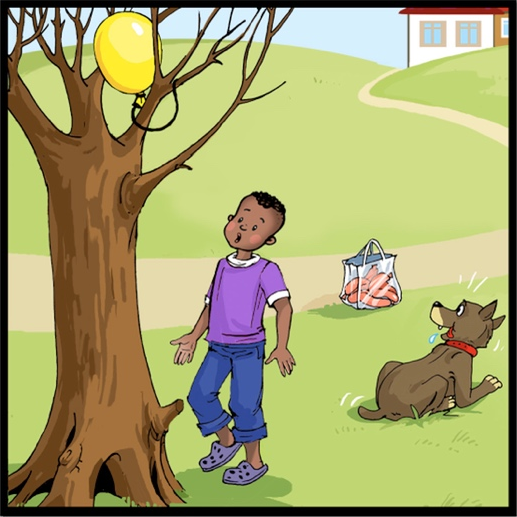}}\hfill
 	\subfloat{\includegraphics[width=0.329\linewidth]{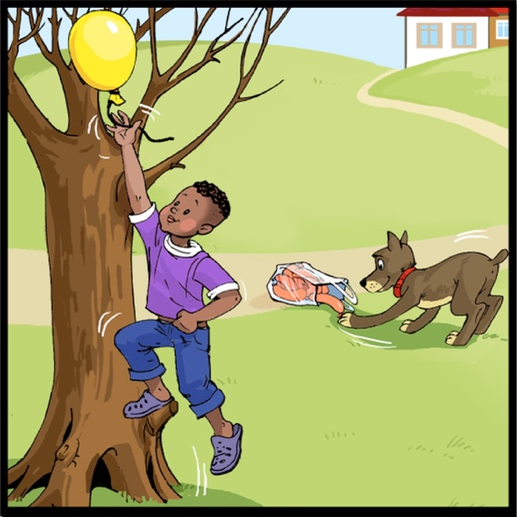}}\hfill	\subfloat{\includegraphics[width=0.329\linewidth]{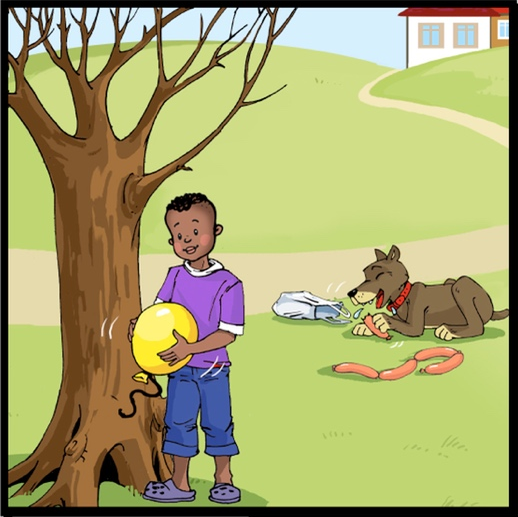}}
	
	\caption{
    A picture sequence used in the MAIN assessment protocol to elicit a narrative from a child.%
 }
	\label{fig:cartoon}
\end{figure}

The first step of a spoken language assessment system~\cite{kaiser_advances_2011,carson_classroom_2013, gillon_better_2019} is to transcribe child speech using an automatic speech recognition~(ASR) system.
This paper describes the development of ASR systems on stories from isiXhosa and Afrikaans preschool children.
ASR for child speech is difficult due to acoustic and linguistic variability~\cite{lee_acoustics_1999, gerosa_acoustic_2007}.
But, compared to previous work, our setting is even more challenging.
First, most studies focus on older children (ages 7 and up)\cite{veeramani_towards_2023, johnson_equitable_2023}, while we are targeting ages 4 and 5; it has been shown that ASR performance deteriorates dramatically for younger children~\cite{yeung_difficulties_2018}.
Second, because many studies focus on early reading ability, few consider spontaneous speech.
Our data consists of spontaneous speech, obtained by prompting a child with a picture sequence (Fig.~\ref{fig:cartoon}).
Third, almost all previous work is on English, with only a handful of publicly available non-English datasets~\cite{claus_survey_2013}.
We consider an extremely low-resource setting where only 5 minutes of transcribed in-domain speech data is available; developing ASR systems without relying on extensive labelled child speech data is crucial for making automated assessment possible in more languages.

In this unique setting, the question is how we should develop our ASR systems.
There is no shortage of strategies, but their benefits are often inconclusive.
For instance, \cite{zhang_improving_2024} reports voice conversion can be a useful data augmentation approach in some settings but not in others.
To give support for or against a particular strategy, more data points are required.
This paper describes a practical case study that provides two data points (Afrikaans and isiXhosa) for confirming or refuting strategies that have been proposed in the past for processing child speech.
Our contribution is therefore not in promoting some new method, but in applying a range of strategies (wider than in most other work) to a particular real-world problem.

We use Whisper~\cite{radford_robust_2023} as a starting point -- a good base model for fine-tuning on transcribed child speech~\cite{fan_benchmarking_2024}. We then consider the following strategies: (1)~parameter efficient fine-tuning~\cite{song_lora-whisper_2024}; (2)~incorporating adult data from the same language to compensate for the absence of child data~\cite{shivakumar_transfer_2018}; (3)~using voice conversion as a data augmentation approach to generate more child-like speech~\cite{shahnawazuddin_voice_2020}; and (4)~semi-supervised learning for labelling untranscribed in-domain child speech that might be available~\cite{lamel_lightly_2002}. 
Of these strategies, we find that using in-domain adult speech -- where an adult produces stories matching our narrative domain -- gives the biggest gain in both Afrikaans and isiXhosa. Semi-supervised learning also helps. 
Voice conversion and parameter-efficient fine-tuning are only beneficial in specific cases.

\section{A Dataset for Oral Narrative Assessment}
\label{sec:setting}

	
	
	

To develop our ASR systems, we use recordings of oral narrative assessments performed on Afrikaans- and isiXhosa-speaking preschool children in South Africa's Western Cape province.
The assessment followed the {Multilingual Assessment Instrument for Narratives} (MAIN) protocol~\cite{gagarina_main_2019}:
a facilitator presents a series of pictures to a child, who is then asked to 
verbalise a story based on the events illustrated in the pictures.
The facilitator then asks the child questions to test their comprehension. 
MAIN was designed to be ecologically valid and culturally neutral, allowing it to assess
children’s narrative production and comprehension skills regardless of linguistic, socioeconomic or cultural backgrounds. 
Fig.~\ref{fig:cartoon} illustrates one of the picture sequences used.
Based on a transcript of the session, a child is then scored using metrics
that assess 
narrative comprehension and the structural complexity of the stories they produce.\footnote{isiXhosa and Afrikaans are two of South Africa's 12 official languages,
with respectively 8M and 7.2M native speakers. isiXhosa is a Southern Bantu language and has a very rich system of agglutinating morphology.
Afrikaans is a West Germanic language that evolved from the 17th-century Dutch spoken by settlers and slaves in the Cape Colony.
Both languages use the Latin alphabet.%
}

In total, 
154 Afrikaans-speaking and 157 isiXhosa-speaking children, aged 4 to 5,
were randomly selected from 55 classrooms to particpate in the study.
Most recordings were sufficiently audible, but some contained noises such as vacuuming, background chatter and microphone fiddling, despite using a directional, noise-cancelling microphone.
We applied minimal filtering since the recordings match the conditions under which our ASR models will operate.
More details on the 
data are given in~\cite{cain2024exploring}.

To prepare the data for ASR, we manually aligned the provided transcripts to the raw recordings using the Praat toolkit~\cite{Praat}.
This resulted in roughly 5 hours of child-speech data for each language.
We refer to this as our in-domain data.
The data for each language was split into a train, development and test set, as shown in Table~\ref{tbl:babaloon_splits}.
To develop ASR models for oral narrative assessments in more languages, collecting and annotating the amount of data presented in Table~\ref{tbl:babaloon_splits} is not feasible.
In our experiments, we therefore explore an extremely resource-limited scenario
{by sampling a random 5-minute subset from the full training set} (last row, Table~\ref{tbl:babaloon_splits}).
We treat this 5-minute training set as the only labelled in-domain child speech for model development.



\begin{table}[t]	
	\mytable
	\caption{Details of the oral narrative dataset, collected from preschool children aged 4 to 5.}
	\begin{tabularx}{1\linewidth}{@{\extracolsep{0pt}}L *{6}{S[table-format=3.0]}}
 
		\toprule
		\multirow{4}{1cm}{} & \multicolumn{3}{c}{Afrikaans} & \multicolumn{3}{c}{isiXhosa} \\
            \cmidrule(l){2-4} \cmidrule(l){5-7}
		Data & {Mins.} & {Spks.} & {Utts.} & {Mins.} & {Spks.} & {Utts.}  \\ 
		\midrule
        Train & 256 & 120 & 5187 & 228 & 118 & 5289  \\
		Development & 47 & 19 & 890 & 40 & 19 & 907  \\
            Test & 30 & 14 & 680 & 35 & 14 & 824 \\
        \addlinespace
		Train 5m & 5 & 5 & 91 & 5 & 5 & 82 \\
		\bottomrule
	\end{tabularx}
	\label{tbl:babaloon_splits}
\end{table}


\section{Base Speech Recognition Model: Whisper}
\label {ssec:model}
We use Whisper~\cite{radford_robust_2023} as our base ASR model
since it gave good performance in previous studies on child ASR~\cite{liu_sparsely_2024, jain_exploring_2024, fan_benchmarking_2024}.
Whisper has five model sizes from which we only consider the small and medium variants due to hardware constrains.
All model sizes were trained on 680k hours of weakly-supervised speech data from 97 languages 
(117k hours are non-English).
The training data
includes 4.1 hours of Afrikaans, but also thousands of hours from related Germanic languages such as Dutch and German. 
On the other hand, no isiXhosa or any closely related languages are included in Whisper. 
{We select Swahili as the language token when fine-tuning on isiXhosa.}

In our experiments, we fine-tune Whisper for up to 5k steps with a batch size of 64 and a learning rate of $\text{1}\cdot\text{10}^{-\text{5}}$.
Without having additional text data to apply an external language model (LM), we benefit from Whisper's implicitly learned LM in final decoding with a beam size of 10.
We apply a length penalty of $-\text{0.5}$ to penalise repetitive hallucinations caused by stuttering and false starts.\footnote{
{
    We also performed development experiments using XLSR~\cite{babu2022xlsr} for ASR. Here we experimented with an external LM trained on the transcriptions of the full training set while using the 5-minute subset for the acoustic model. The LM provided benefits with XLSR, but the absolute performance was always worse than using Whisper with its implicit LM learned in its decoder.  %
    }%
}

\section{Experiments}
\label{sec:experiments}

We explore different strategies to build and improve our ASR models.
While some of these have been effective {in child-speech ASR in prior work,}
we now evaluate them in our own unique low-resource {scenario}.
{Before considering each strategy in turn, we}
look at the results of our base models, which are trained by fine-tuning Whisper (small and medium) using 
{5 minutes} 
of {transcribed} in-domain child speech (last row, 
Table~\ref{tbl:babaloon_splits}).

The {word error rates (WERs)} 
for these models are presented in the first row of Table~\ref{tbl:baseline}.
Despite using the same amount of training data, the WER for Afrikaans is significantly better than for isiXhosa: 47.4\% compared to 80.4\% for Whisper medium.
This can partly be attributed to the mismatch between isiXhosa and the languages included in Whisper's pretraining data.
Moreover, a previous study on adult ASR~\cite{barnard_nchlt_2014} found that it is more challenging for ASR models to transcribe agglutinative languages like isiXhosa than Germanic languages like Afrikaans.
The character error rate (CER) for these 5-minute models are 27.9\% for Afrikaans and 34.0\% for isiXhosa (not shown in the table); this smaller difference in error rate shows that isiXhosa is more severely penalised for inaccurate word predictions.

To situate the results that follow, we consider lower and upper bounds.
Representing the case where no child speech data is available, we apply Whisper without any fine-tuning.
This baseline produces nonsensical results for isiXhosa with a WER of 105.6\%, whereas for Afrikaans the results are poor but not random with a WER of 88.1\% (third row, Table~\ref{tbl:baseline}, Whisper medium).
Topline results (when fine-tuning on the full training sets) are shown in the last row of Table~\ref{tbl:baseline}.

We now consider strategies to improve {our 5-minute models.}



\begin{table}[t]	
	\mytable
	\caption{WERs (\%) on development data for Afrikaans and isiXhosa, trained on in-domain child speech.}
	
	\begin{tabularx}{1\linewidth}{@{\extracolsep{0pt}}L *{4}{S[table-format=2.1]}}
		\toprule
		\multirow{4}{1cm}{} & \multicolumn{2}{c}{Afrikaans} & \multicolumn{2}{c}{isiXhosa} \\
            \cmidrule(l){2-3} \cmidrule(l){4-5}
		Model & {small} & {medium} & {small} & {medium} \\ [-0.3mm]
		\midrule
		5m-child  & 54.4 & 47.4 & \textbf{86.0} & \textbf{80.4} \\
		5m-child (LoRA) & \textbf{52.1} & \textbf{41.1} & 92.4 & 88.7 \\
        \addlinespace
		Baseline: Whisper w/o fine-tuning & 90.3 & 88.1 & 107.7 & 105.6 \\
		Topline & 21.3 & 18.3 & 54.4 & 50.9 \\
		\bottomrule
	\end{tabularx}
	\label{tbl:baseline}
\end{table}

\subsection{Does parameter-efficient fine-tuning help?}

\textbf{Related work:}
Low-rank adaptation (LoRA)
is a method where a network is efficiently updated without 
having to retrain all the network parameters~\cite{hu_lora_2022}.
Apart from computational efficiency, LoRA has also resulted in better performance by preventing overfitting~\cite{song_lora-whisper_2024}. 
E.g.~\cite{liu_sparsely_2024} achieved better child-speech ASR performance with LoRA fine-tuning than updating all weights.
In contrast, 
in~\cite{fan_benchmarking_2024} the authors fine-tune Whisper for English child speech and find that LoRA is not always beneficial.
Given these inconsistencies, we provide two additional data points to validate the effectiveness of LoRA for child-speech ASR.


\textbf{Setup:} 
{LoRA inserts learnable decomposition matrices while freezing the original model parameters. We specifically add}
low-rank matrices with $r=\text{32}$ to {the} attention layers $\{\mathbf{W}_{q}, \mathbf{W}_{\text{k}}, \mathbf{W}_{\text{v}}, \mathbf{W}_{\text{out}}\}$ {with a learning rate of $\text{1}\cdot\text{10}^{-\text{4}}$.}
We fine-tune using the {5-minute} in-domain sets.

\textbf{Results:} Comparing the LoRA fine-tuned variants (second row, Table~\ref{tbl:baseline}) to the full-parameter fine-tuned models (first row), 
we see that for Afrikaans {LoRA}
helps by improving WER from 47.4\% to 41.1\%  but it hurts on isiXhosa with WER going from 80.4\% to 88.7\%.
{This shows that LoRA is not always beneficial.}
{It also highlights again that we might need particular strategies when dealing with underrepresented languages like isiXhosa -- even though we are in the same domain, the absolute performance and the benefits of particular strategies differ dramatically between Afrikaans and isiXhosa.}

Based on these results, we proceed with Whisper medium  but without LoRA in the experiments below.



\subsection{Does out-of-domain adult speech help?}
\label{ssec:ood_adult}
\textbf{Related work:}
{Annotated child speech is very limited in most languages.}
To compensate for this, 
{one approach is to use transcribed}
adult speech from the target language to combine with child speech during training.
While some found this to be fruitful~\cite{shivakumar_transfer_2018}, others have not seen any benefit~\cite{elenius_adaptation_2005, fainberg_improving_2016}.
We explore the effectiveness of this strategy
using 
adult speech from {the same language but} a different domain {than oral narratives}.


\textbf{Setup:} 
We sample 
out-of-domain adult speech from the NCHLT speech corpus~\cite{barnard_nchlt_2014}.
There
are roughly 56 hours of speech from around 200 speakers
for Afrikaans and isiXhosa {each}.
{The audio is mostly read text from government documents.}
We incorporate the adult data using a similar transfer learning approach to~\cite{shivakumar_transfer_2018}: we first fine-tune Whisper using out-of-domain adult data, and then further fine-tune using the 5-minute,  in-domain child data.

\textbf{Results:}
WERs are presented in the upper section of Table~\ref{tbl:adult_in_domain}.\footnote{Results repeating from a previous table are marked with an asterisk~($\ast$).}
First, we look at the model {fine-tuned} only 
on the adult data (NCHLT-adult){, achieving 94.5\% and 94.3\% WERs on Afrikaans and isiXhosa, respectively. These are poor scores and provide little or no benefit over the baseline models in Table~\ref{tbl:baseline} (88.1\% and 105.6\%).} 
This aligns with the consensus that adding unmodified adult speech from a different domain does not help~\cite{elenius_adaptation_2005, fainberg_improving_2016, bell2020adaptation}.
Where we do see a benefit is when the model is further fine-tuned on the in-domain child data (NCHLT-adult $\rightarrow$ 5m-child),  with the WER improving by 4.8\% {absolute} for Afrikaans and 6.5\% for isiXhosa over the exclusive 5m-child model.
The larger boost for isiXhosa could be ascribed to Whisper being exposed to isiXhosa for the first time. 
Since isiXhosa is underrepresented in Whisper, we also trained a multilingual Whisper model including languages related to isiXhosa, followed by 5-minute in-domain 
fine-tuning.\footnote{We use four Nguni languages {(isiNdebele, Siswati, isiZulu and also isiXhosa)} from the NCHLT corpus, amounting to 200 hours of training data.} But this yielded only a 1.3\% absolute improvement in WER {(NCHLT-adult-multi. $\rightarrow$ 5m-child, Table~\ref{tbl:adult_in_domain}).}

Our findings {therefore} indicate that using adult speech from a different domain can be beneficial when working with very limited amounts of target domain data (in the order of a few minutes).
We found, however, that this gain did not realise when more in-domain training data is available:
the topline results in Table~\ref{tbl:baseline} did not improve by including adult speech.

\begin{table}[!t]	
	\mytable
	\caption{
    {WERs (\%) on development data when in- and out-of-domain transcribed adult data is used to supplement child speech.}
    }
	\begin{tabularx}{1\linewidth}{@{\extracolsep{4pt}}L*{2}{S[table-format=2.1]}}
		\toprule
		
		Model & {Afrikaans} & {isiXhosa} \\
		\midrule
		\underline{\textit{Out-of-domain adult speech:}} & & \\
		5m-child & 47.4$^\ast$ & 80.4$^\ast$ \\
		NCHLT-adult & 94.5 & 94.3 \\
		NCHLT-adult $\rightarrow$  5m-child & \textbf{42.6} & 73.9 \\
		NCHLT-adult-multi. $\rightarrow$ 5m-child & {---} & \textbf{72.6} \\[3pt]

		\underline{\textit{In-domain adult speech:}} & & \\
		5m-child  & 47.4$^\ast$  & 80.4$^\ast$ \\
		5m-child + 5m-adult & 43.1 & 78.4 \\
		5m-child + 30m-adult & \textbf{33.7} & \textbf{70.2} \\
		\bottomrule
	\end{tabularx}
	\label{tbl:adult_in_domain}
\end{table}

\subsection{Does in-domain adult speech help?}
\label{ssec:id_adult}

\textbf{Related work:}
In the previous experiment, we (and others) showed that incorporating adult speech from a different domain can be effective despite the linguistic mismatch to the target domain.
However, these improvements are not as substantial as those achieved with even a small amount of in-domain child data, e.g., comparing 5m-child to NCHLT-adult $\rightarrow$ 5m-child in Table~\ref{tbl:adult_in_domain}.
This prompted us to ask whether a model {can} benefit from recordings of an adult speaker that matches the target domain of child oral narratives.
This would be much easier to obtain than capturing child speech data (which brings major practical and ethical challenges).


\textbf{Setup:}
We asked
an adult 
speaker from each target language to record themselves reading transcriptions from the in-domain training set which were not included in our 5-minute training set.
For each language, we {collected} 30 minutes of adult speech.
We also {sampled} a smaller set of 5 minutes {of adult audio, for comparison.}
We pool the child (in-domain) and adult (in-domain) data for fine-tuning, {rather than training sequentially as in Sec.~\ref{ssec:ood_adult}}.

\textbf{Results:}
{WERs are}
shown in the bottom section of Table~\ref{tbl:adult_in_domain}.
For Afrikaans and isiXhosa, adding 5 minutes of in-domain adult data (5m-child + 5m-adult) improves WER by 4.3\% and 2.0\%, respectively.
For Afrikaans, using only 5 minutes of in-domain adult data nearly matches the performance of using 52 hours of out-of-domain adult data (NCHLT-adult $\rightarrow$ 5m-child). 
When including up to 30 minutes of in-domain adult data,
{we achieve the best results that we have seen so far for both languages (33.7\% and 70.2\%).}

Adult in-domain data is beneficial. Our analysis is idealised in that we assume we have text corresponding to child oral narratives -- but these will still be easier to obtain than child recordings.


\subsection{Does voice conversion help?}
\label{ssec:vc}
\textbf{Related work:} 
Secs.~\ref{ssec:ood_adult} and~\ref{ssec:id_adult} showed that both in- and out-of-domain adult speech helps.
However, the variability and pronunciation of child speech, especially at the age we consider, are vastly different from those of adults~\cite{lee_acoustics_1999, gerosa_acoustic_2007}.
To address this acoustic mismatch, 
voice conversion (VC) has proven effective by generating 
child-like speech from adult speakers~\cite{shahnawazuddin_voice_2020, singh_data_2021, shuyang2023data, zhang_improving_2024}.
But {improvements are} not always consistent across domains and tasks~\cite{zhang_improving_2024}.


\textbf{Setup:}
{We need target child speech to serve as reference for a VC system. To mimic the realistic scenario where we do not have appropriate child speech in the target language upfront,}
we opt to use clean child speech from a different language
-- a form of cross-lingual VC~\cite{baas2023voicea}.
Specifically, we use 10 hours of read speech from 40 British English children. 
We apply VC to both our in- and out-of-domain 
adult speech 
using the recent kNN-VC system, a light-weight method giving state-of-the-art results~\cite{baas2023voiceb}.

\textbf{Results:}
{WERs}
are shown in Table~\ref{tbl:vc}.
Training a model on the converted out-of-domain adult speech and then 
further
fine-tuning on in-domain child speech (NCHLT-adult-vc $\rightarrow$ 5m-child) shows no 
gain compared to using the unmodified adult speech (NCHLT-adult $\rightarrow$ 5m-child).
{However, when in-domain adult speech is converted, we see consistent improvements of between 2 and 3\% in WER, with the second-to-last  row in Table~\ref{tbl:vc} giving the best results achieved so far.}
{Our findings differ from~\cite{shahnawazuddin_voice_2020, singh_data_2021}, where} converted adult speech from a different domain {helped, but it matches~\cite{zhang_improving_2024}, which showed that improvements from VC can be setting-specific.}
{We show here specifically that it does not help to convert data with a large linguistic mismatch from the target domain (out-of-domain adult speech), but when} the linguistic mismatch {disappears} 
(in-domain adult speech){, then VC can be used to compensate for the acoustic mismatch.}

{The improvements we achieved here and in Sec.~\ref{ssec:id_adult} comes mostly from better linguistic rather than acoustic modelling.
To show this, we also did a data augmentation experiment where we took the 5-minute child data and converted it to voices of other children.
This increases acoustic diversity, but it gave no performance improvement.
(This is different from~\cite{zhang_improving_2024}, but~\cite{baas2022voice} showed that the benefits of VC-based data augmentation is very dependent on the amount of training data.)}




\begin{table}[!t]
	\mytable
	\caption{WERs (\%) on development data when VC is applied to in- and out-of-domain adult speech to get more child-like speech.}
	\begin{tabularx}{1\linewidth}{@{\extracolsep{4pt}}L*{2}{S[table-format=2.1]}}
		\toprule
		Model & Afrikaans & isiXhosa \\
		\midrule
		\underline{\textit{Out-of-domain}:} &&\\
		NCHLT-adult $\rightarrow$  5m-child & 42.6$^\ast$ & \textbf{73.9}$^\ast$ \\
		NCHLT-adult-vc $\rightarrow$ 5m-child  & \textbf{42.3} & 74.7 \\[3pt]
		\underline{\textit{In-domain}:} &&\\
		5m-child + 5m-adult & 43.1$^\ast$ & 78.4$^\ast$ \\
		5m-child + 5m-adult + 5m-adult-vc & 41.1 & 75.7 \\
		5m-child + 30m-adult & 33.7$^\ast$ & 70.2$^\ast$ \\
		5m-child + 30m-adult + 30m-adult-vc & \textbf{31.6} & \textbf{68.1} \\
        30m-adult + 30m-adult-vc & 53.9 & 76.1 \\
		\bottomrule
	\end{tabularx}
	\label{tbl:vc}
\end{table}

\begin{table}[t]	
	\mytable
	\caption{WERs (\%) on development data when semi-supervised learning is used to get additional training data from unlabelled child speech.}
	\begin{tabularx}{1\linewidth}{@{\extracolsep{4pt}}L*{2}{S[table-format=2.1]}}
		\toprule
		Model & Afrikaans & isiXhosa \\
		\midrule
		5m-child  & 47.4$^\ast$  & 80.4$^\ast$ \\
  		5m-child $\rightarrow$ 5m-child + 30m-child-ss & \textbf{39.6} & 76.5 \\
		  5m-child $\rightarrow$ 5m-child + 1h-child-ss & 39.9 & \textbf{74.5} \\
		5m-child $\rightarrow$ 5m-child + 2h-child-ss & 41.3 & 75.8 \\
		\bottomrule
	\end{tabularx}
	\label{tbl:semi_supervised}
\end{table}

\subsection{Does semi-supervised learning help?}
\label{ssec:semi}
\textbf{Related work:}
An ASR model trained on limited amounts of data can be used to generate pseudo labels for unlabelled audio~\cite{thomas_deep_2013, weninger_semi_2020, wallington_learning_2021}.
This strategy, a form of semi-supervised learning, can help improve a model by increasing the amount of training data without having perfect transcriptions.
This proved helpful for child-speech ASR in~\cite{wang_low_2021}.
It is also very relevant since getting more unlabelled audio is often much easier than getting transcriptions.

\textbf{Setup:}
We apply the 5-minute fine-tuned ASR models (first row,  Table~\ref{tbl:baseline}) to the remaining in-domain training data (first row, Table~\ref{tbl:babaloon_splits}), which we treat as 
unlabelled audio from the target domain.
The resulting pseudo labels are combined with the transcribed 5-minute set to train a new model.
Using all the predicted labels without filtering can hurt performance~\cite{lamel_lightly_2002}.
We {therefore} implement a data quality filtering strategy~\cite{wang_low_2021}: 
we 
{only keep the pseudo labels with the highest log-likelihood scores according to the Whisper model.}
{We consider the top}
30-minute, 1-hour and 2-hour 
{predictions.}

\textbf{Results:}
{WERs}
are shown in Table~\ref{tbl:semi_supervised}.
Both languages benefit from being exposed to more in-domain data even without perfect transcriptions. 
The 5.9\% WER improvement for isiXhosa is 
{noteworthy} 
considering that the model used to generate the transcriptions has a WER of 80.4\%.
{For both languages, the biggest improvement comes from using the the top 1 hour of predicted transcriptions,}
after which performance starts to 
{deteriorate.}

\subsection{Combined systems}

We now combine all the best strategies.
We fine-tune
Whisper on out-of-domain adult data (Sec.~\ref{ssec:ood_adult}) and then further fine-tune on the pooled 5-minute in-domain child, unmodified in-domain adult~(Sec.~\ref{ssec:id_adult}), and in-domain adult converted~(Sec.~\ref{ssec:vc}) data. 
We then use this model to predict transcripts for the remaining unlabelled in-domain child speech~(Sec.~\ref{ssec:semi}), add these predictions to the training pool, and fine-tune for a third time.
The results
for this 
model are presented in Table~\ref{tbl:combined}.
It achieves the
best WERs 
on the development data: 29.8\% on Afrikaans and 62.1\% on isiXhosa.
We also, for the first time, show WERs on our held-out test set~(third row, Table~\ref{tbl:babaloon_splits}). These correlate well with the scores on the development sets:
{on the test data, we respectively achieve a 13.8\% and 21.4\% absolute improvement for Afrikaans and isiXhosa over the base model trained only on the 5-minute sets.}
This combined best system (utilising only 5 minutes of labelled data) also comes within 12\% of the topline system (utilising more than 3 hours of labelled in-domain data).
For reference, the CER of our best 5-minute combined system is 17.3\% and 17.4\% on the Afrikaans development and test sets, respectively, and 26.1\% and 22.6\% on the isiXhosa sets.



\begin{table}[!t]	
	\mytable
	\caption{WERs (\%) on development and test data when combining the best strategies.}
	\begin{tabularx}{1\linewidth}{@{\extracolsep{0pt}}L*{4}{S[table-format=2.1]}}
		\toprule
		\multirow{4}{1cm}{} & \multicolumn{2}{c}{Afrikaans} & \multicolumn{2}{c}{isiXhosa} \\
            \cmidrule(l){2-3} \cmidrule(l){4-5}
            Model &{dev}&{test}&{dev}&{test} \\
            \midrule
            Base: 5m-child  &  47.4$^\ast$  & 42.8 & 80.4$^\ast$  & 78.7 \\
            
            \addlinespace

		Combined: NCHLT-adult \\ $\rightarrow$ 5m-child + 30m-adult + 30m-adult-vc \\ $\rightarrow$  [same as above] + 1h-child-ss & 29.8 & 29.0 & 62.1 & 57.3 \\
        \addlinespace

  		Topline & 18.3$^\ast$  & 17.5 & 50.9$^\ast$  & 51.9\\

		\bottomrule
	\end{tabularx}
	\label{tbl:combined}
\end{table}

\section{Conclusion}



We looked at strategies previously proposed for child ASR, but in a wider range of combinations for the unique setting of recognising oral narratives from children aged 4 to 5.
Using only 5 minutes of in-domain data we found the following: Using out-of-domain adult data is more beneficial on isiXhosa than on Afrikaans, since isiXhosa is under-represented in the Whisper base model. Parameter-efficient fine-tuning is also not consistent on the two languages.
Voice conversion helps, but only if it is applied to in-domain data.
Semi-supervised learning helps on both languages.
A combined system shows that the best strategies are complementary when applied together.

The ASR models developed here represent the first step towards the larger goal of automating oral narrative assessments. 
In the full assessment system, predicted transcripts will be fed into a subsequent model to score a child's narrative ability. 
Although our WERs are still high (including our topline results), previous studies have shown that insights can still be extracted from noisy transcripts despite high WERs~\cite{tao_audio_2019, pugh_saywhat_2021}.
Future work will investigate this in a full oral narrative assessment system.


\clearpage
\balance{}
\bibliographystyle{IEEEtran}
\bibliography{refs}

\end{document}